\documentclass[aps,pra,twocolumn,showpacs,amsmath,amssymb,preprintnumbers,superscriptaddress,10pt]{revtex4-1}

\usepackage{graphicx}
\graphicspath{{figures/}}
\usepackage[bookmarks=false]{hyperref}
\usepackage{color}
\usepackage[capitalise]{cleveref}
\crefname{section}{Sec.}{Secs.}
\Crefname{section}{Section}{Sections}

\definecolor{pink}{RGB}{255,0,255}

\begin{document}

\title{Comment on ``Inherent security of phase coding quantum key distribution systems against detector blinding attacks'' [Laser Phys.\ Lett.\ 15, 095203 (2018)]}

\author{Aleksey~Fedorov}
\affiliation{Russian Quantum Center, Skolkovo, Moscow 143025, Russia}
\affiliation{QRate, Skolkovo, Moscow 143025, Russia}
\affiliation{QApp, Skolkovo, Moscow 143025, Russia}

\author{Ilja~Gerhardt}
\affiliation{3.~Institute of Physics, University of Stuttgart and Institute for Quantum Science and Technology, Pfaffenwaldring 57, D-70569 Stuttgart, Germany}
\affiliation{Max Planck Institute for Solid State Research, Heisenbergstra\ss e 1, D-70569 Stuttgart, Germany}

\author{Anqi~Huang}
\affiliation{Institute for Quantum Information \& State Key Laboratory of High Performance Computing, College of Computer, National University of Defense Technology, Changsha 410073, People's Republic of China}

\author{Jonathan~Jogenfors}
\affiliation{Department of Electrical Engineering, Link{\" o}ping University, SE-58183 Link{\" o}ping, Sweden}

\author{Yury~Kurochkin}
\affiliation{Russian Quantum Center, Skolkovo, Moscow 143025, Russia}
\affiliation{QRate, Skolkovo, Moscow 143025, Russia}

\author{Ant{\'i}a~Lamas-Linares}
\affiliation{Texas Advanced Computing Center, The University of Texas at Austin, Austin, Texas, USA}

\author{Jan-{\AA}ke~Larsson}
\affiliation{Department of Electrical Engineering, Link{\" o}ping University, SE-58183 Link{\" o}ping, Sweden}

\author{Gerd~Leuchs}
\affiliation{Max Planck Institute for the Science of Light and University of Erlangen-N{\" u}rnberg, D-91058 Erlangen, Germany}

\author{Lars~Lydersen}
\affiliation{Kringsj{\r a}vegen 3E, NO-7032 Trondheim, Norway}

\author{Vadim~Makarov}
\email{makarov@vad1.com}
\affiliation{Russian Quantum Center, Skolkovo, Moscow 143025, Russia}
\affiliation{National University of Science and Technology MISIS, Moscow 119049, Russia}

\author{Johannes~Skaar}
\affiliation{Department of Technology Systems, University of Oslo, Box~70, NO-2027 Kjeller, Norway}

\date{September~11,~2018}

\maketitle

In Ref.~\onlinecite{balygin2018}, Balygin and his coworkers consider a faked-state attack with detector blinding on Bennett-Brassard 1984 (BB84) quantum key distribution (QKD) protocol. They propose a countermeasure to this attack in a phase-coded system that watches for an abnormally low number of detections in the outer time slots 1 and 3. If
the eavesdropper does not pay attention to the outer time slots, the countermeasure will reveal that the attack is being performed (see Secs.~6, 7, and Fig.~1(b) in Ref.~\onlinecite{balygin2018}). This approach is conceptually similar to earlier work on non-blinding attacks \cite{ferreiradasilva2012}.

However, in the faked-state attack \cite{makarov2005} the eavesdropper Eve uses a replica of Bob's setup to detect all quantum states emitted by Alice, then induces her exact detection results in Bob's apparatus. Since Eve is using a replica of Bob's setup, she would register detections in the outer time slots, then induce the same detection results in Bob's apparatus by resending additional bright light pulses centered in the time slots 1 and/or 3. Note that Eve will occasionally register a double click, i.e.,\ simultaneous detection events in both her detectors caused by dark counts or multiphoton pulses from Alice. She may also in some implementations register multiple clicks in adjacent time slots. She might induce such multiple clicks in Bob using faked states similar to those constructed for distributed-phase-reference protocols \cite{lydersen2011}. I.e.,\ Eve might even replicate imperfections such as double clicks and dark counts that would exist in Bob's equipment. {\em This would mean that Bob's detection events are exactly the events measured by a copy of Bob's setup (conditioned on Bob's basis choice), and are therefore indistinguishable from the detection events without the attack. The statistics of these detections at Bob would thus be indistinguishable from the statistics without the attack, and the countermeasure is ineffective.}

Although the search for technical countermeasures against the attacks on detectors continues \cite{huang2016,sajeed2016,maroy2017,koehler-sidki2018,koehler-sidki2018a}, so far the only practical scheme proven to be immune against these attacks is measurement-device-independent QKD \cite{lo2012,tang2016}.

We finally make a minor remark that Ref.~\onlinecite{balygin2018} uses a simplified model of the blinded detector with a single threshold $P_\text{th}$ at which it begins to make clicks with a non-zero probability. In actuality, the click probability increases gradually at powers higher than that, and there is another threshold $P_\text{always} > P_\text{th}$ at which it becomes unity \cite{lydersen2010a,huang2016}. Although this detail is inconsequential for the argument presented in Ref.~\onlinecite{balygin2018}, it will have to be heeded when constructing the actual attack.

\end{document}